\documentstyle[12pt]{article}

\newcommand{\ketss}[3]{\left| #1 \right\rangle_{#2}^{#3}}
\begin{document}

\begin{tabbing}
\hskip 11.5 cm \= {Imperial/TP/93-94/56}
\\
\> July 1994  \\
\> gr-qc/9408023 \\
\> Minor revisions 10/94\\
\end{tabbing}
\vskip 1cm

\begin{flushleft}
{\Large \bf The Theory of Everything vs  the Theory of
Anything\footnote{To appear in the proceedings of ``The Birth of the
Universe and Fundamental Forces'', Rome, May 1994,  F. Occhionero
Editor, Springer Verlag}
}\\
\bigskip
{\it Andreas Albrecht}\\
\bigskip
Blackett Laboratory, Imperial College\\
Prince Consort Road, London SW7 2BZ  U.K.
\end{flushleft}

%\title{The Theory of Everything vs  the Theory of Anything}
%\bigskip
%\author{Andreas Albrecht}
%\bigskip
%\institute{Blackett Laboratory, Imperial College\\
%Prince Consort Road, London SW7 2BZ  U.K.}

%\maketitle

\section{Introduction}
It is often stated that physicists are concerned with producing a
``Theory of Everything''.  However, it is clear that we shall never probe
all possible interactions at all possible  energies.  By our very
nature we are limited to exploring a very special part of superspace,
where  spacetime behaves nice and classically, and where matter is
distributed in a very orderly way.  Indeed we benefit greatly from these
special properties and (at least to some extent) could not survive
without them.
To what extent can  our limited set of observations  be used to pin
down the specifics of a
``Theory of Everything''?  In the limit where the links are
arbitrarily tenuous, a ``Theory of Everything'' might become a
``Theory of Anything''.
A clear understanding of what we can and can not expect to learn about
the universe is particularly  important in the field of particle
cosmology.   The aim of this article is to draw attention to some key
issues which arise in this context, in the hopes of fostering further
discussion.

In Sect. 2 I explore the features of our own observations with
suggest that a ``Theory of Anything'' might be the best we could hope
for.  I give particular emphasis to the ``choice of clock'' issue
which seems to have been under-emphasized in this context.

In Sect. 3 I discuss the important role played by posing {\em
conditional}  probability questions when dealing with probability
distributions rather than idealized points in classical phase space.  I
emphasize the need to stay  as far away  from ``anthropic'' arguments
as possible,  but point out that  quantum mechanics guarantees that we
can never achieve the idealized classical view.

Section 4  explores possible  implications of these ideas on
inflationary cosmology.  The idea that a variety of different inflaton
potentials may contribute to worlds ``like ours''
might give a physical measure of what is ``natural''
for an inflation potential
which is quite different than those previously used.  I speculate on a
possible outcome of this line of reasoning which is particularly
interesting, and which
removes some of the least attractive features of the
inflationary cosmology.

Section 5 gives my conclusions.

%\pagebreak

\section{The inaccessability of  a ``Theory of Everything''}

Suppose there really is a ``Theory of Everything'' (TOE) which
accurately describes nature at the most fundamental level.  Would
we be able to tell that this was the case based on accessible
experiments?
Here are some arguments which suggest that specific tests of such a
theory would elude us.

\subsection{Multiple Vacua}

Many candidates for the TOE (for example superstring theory) have more
than one stable (or ultra
long-lived) vacuum state.  Many of these states have different
symmetries, or even different spacetime dimensions compared with the
universe we observe. The non-commutivity of variables and their
conjugate momenta ensure that no quantum state can represent a single
point in classical phase space, and thus non-zero probabilities are
assigned  to a range of possible states
Any cosmological history (starting with the
high energies and densities of the big bang) would certainly assign
non-trivial probabilities to the alternative vacua.  Our
observations would necessarily be limited to ``small'' excitations
around a single vacuum state.  Thus from our point of view the TOE
would not predict the laws of physics we observe, but simply tell us
that the laws we observe were chosen from a wide range of
possibilities.

\subsection{Renormalization of Couplings}

One interesting point which has come up in the context of wormhole
physics is that fluctuations in the  gravitational field may
renormalize the coupling constants relevant to processes we
observe. In many cases the result is a continuum of non-interacting
superselection sectors, each with different values for the low-energy
couplings\cite{c88}.  Again, the consequence is that the physical laws we
observe are not specified by the TOE,  but chosen from many possibilities.

%\pagestyle{myheadings}
%\markboth{ Andreas Albrecht}{The Theory of Everything vs the Theory of
%Anything}

\subsection{Choice of Clock}

The next two subsections are a bit outside the main flow of this
article.  The discussion is somewhat more technical, and I am taking
the opportunity to give wider publicity to what I feel is an important
ambiguity which  arises in quantum cosmology.  The reader who is
willing to simply accept the premise that a TOE can leave a great
deal of ambiguity regarding the physical laws we actually observe may
wish to skip the next two subsections.

In a time reparameterization invariant theory, (such as a relativistic
point particle or general relativity) the starting point for the
quantum theory is a superspace with no time defined.  A particular
degree of freedom (or subsystem) is then assigned the role of a clock,
and the evolution  of the rest of the physical world as a function of
time is then evaluated by inspecting the correlations between the clock
subsystem and the rest of the world.  (Although  such a construction
may seem peculiar to physicists who are used to treating time as some
external parameter, the procedure just described gives a
good operational definition of time as it is actually used in a
laboratory.)

To sketch this construction consider a superspace
$S$, with a measure defined so that given a
state $|\psi\rangle_S$ in $S$ assigns the
probability $|{ }_S\!\langle \psi | X\rangle_S |^2$ to
state $|X\rangle_S|^2$.  Consider a clock
subsystem $C$ defined by the partition
\begin{equation}
S =  C\otimes R
\label{CRpartition}
\end{equation}
with $R$  representing the ``rest'' of the superspace.  Part of the
construction involves defining the special basis for the $C$ subspace
which gives the time eigenstates of the clock. I denote this basis
\begin{equation}
\left\{ \ketss{t_i}{C}{ }\right\}.
\label{clockbasis}
\end{equation}
I have no  qualms  about labeling time (or anything else usually
described by a continuous variable)
using  a discrete index (such as the index $i$ labeling clock
eigenstates in Eq \ref{clockbasis}). The difference between a
continuum and an arbitrarily closely spaced discrete label can not
possibly be physically  observable.

A basis $\left\{\ketss{j}{R}{}\right\}$ spanning the $R$ space can also be
chosen, and
together with the
clock basis a tensor product basis spanning $S$ can be constructed.
Thus a state $\ketss{\psi}{S}{}$ in superspace can be written
\begin{equation}
\ketss{\psi}{S}{} =
\sum_{ij}\alpha_{ij}\ketss{t_i}{C}{}\ketss{j}{R}{}.
\label{CRexpand}
\end{equation}
Defining
\begin{equation}
\ketss{\phi_i}{R}{} \equiv \sum_j\alpha_{ij}\ketss{j}{R}{}
\label{phidef}
\end{equation}
gives
\begin{equation}
\ketss{\psi}{S}{} = \sum_i \ketss{t_i}{C}{}\ketss{\phi_i}{R}{}.
\label{tphiexpand}
\end{equation}
The state $\ketss{\psi(t_i)}{R}{}$ of subsystem $R$ at time $t_i$ is
determined by conditioning
(projecting) on clock state $\ketss{t_i}{C}{}$,  giving
\begin{equation}
\ketss{\psi(t_i)}{R}{} = \ketss{\phi_i}{R}{}.
\label{psioft}
\end{equation}
Equation \ref{tphiexpand} tells us that ``when the clock is in state
$\ketss{t_i}{C}{}$ the $R$ subsystem is in state
$\ketss{\phi_i}{R}{}$ (with unit probability)''.

The physics of this system is contained entirely in the
$\alpha_{ij}$'s.  From these one can deduce  the initial state
$\ketss{\psi(t_0)}{R}{}$ and the subsequent time evolution
$\ketss{\psi(t_i)}{R}{}$.  A different
initial state evolving in a different manner would correspond to a
different set of $\alpha_{ij}$'s.

However, a different set of $\alpha_{ij}$'s just corresponds to a
different choice of clock.  To see this, construct a unique integer
label $k(i,j)$ for each pair $i,j$.  One can then write
\begin{equation}
\ketss{k}{S}{} \equiv \ketss{i(k)}{C}{}\ketss{j(k)}{R}{}
\label{kSdef}
\end{equation}
and
\begin{equation}
\ketss{\psi}{S}{} = \sum_k \alpha_k \ketss{k}{S}{}.
\label{kSexpand}
\end{equation}
If one expands $\ketss{\psi}{S}{}$ in a  different basis
$\ketss{k'}{S}{}$ one will get a {\em different} set of expansion
coefficients ${\tilde \alpha}_{k'}$.
If one then identifies elements of this
new basis with tensor product states labeled by $i'(k')$ and $j'(k')$
one gets
\begin{equation}
\ketss{\psi}{S}{} =
\sum_{i'j'}{\tilde \alpha}_{k'(i',j')}\ketss{i'}{C'}{}\ketss{j'}{R'}{}.
\label{ijprimeexpand}
\end{equation}
Note that this amounts to the construction of a {\em new} partition
$S = C'\otimes R'$ which in general will have nothing to do with $S=
C\otimes R$.

Since bases can be chosen for $S$ which results in arbitrary
$\tilde \alpha_{k'}$'s, partitions of the form $S=C'\otimes R'$ can be
constructed which correspond to all possible $\tilde \alpha_{i'j'}$'s.
Thus, starting with $\ketss{\psi}{S}{}$ one can produce all possible
states evolving according to all possible time evolutions simply by
choosing a suitable clock.

\subsection{Choice of Clocks: Further Discussion}

A random choice of $\alpha_{ij}$'s  will generally not give a sensible
time evolution.  In the general case $\ketss{\psi(t_i)}{R}{}$ and
$\ketss{\psi(t_{i+1})}{R}{}$ will not be  related to one another by a
simple Schr\"{o}dinger equation. (Or, putting it another way, the
Hamiltonian will have some totally random time dependence.)  Thus
most partitions of $S$ into $C\otimes R$  will not identify good clock
subsystems.

One of the important features of our universe is that there
are many subsystems  which can serve equally well as clocks (which is
why there can be such a thing as a clock factory, for example).  That
means that if one sets up a theory which accurately describes our
world, there is large class of partitions $S=C\otimes R$ which
simply correspond to switching from one clock to another, all of which
are essentially describing the {\em same} initial conditions with the
same time evolution.  This fact does not conflict with the statement
that there are {\em other} choices of the $Clock\otimes Rest$ partition
which correspond to something totally different.

Where does the Wheeler DeWitt  equation fit into my discussion? Under
certain circumstances the the Hilbert space can be further partitioned
into fields representing matter and gravitational degrees of
freedom. If the whole system obeys Einstein gravity, one of the
required constraint equations is the Wheeler DeWitt equation.
Since any physical system (and many unphysical ones!) can be described
by a suitable choice of partition $S=C\otimes R$,  some will obey a
suitable Wheeler DeWitt equation and many will not.  Of those that do,
all sorts of boundary conditions might be satisfied, and the boundary
conditions would not be uniquely specified by $|\psi\rangle_S$.

The  ``problem of time'' has been studied a great deal in the context
of quantum gravity\cite{isham93,kuchar92}. (See for example
\cite{rovelli90,rovelli&smolin94,onder&tucker94} for work closely
related to this article.) My discussion embraces the
``Page-Wooters'' (PW) picture of time.  The ambiguity in the
dynamics which I discuss here has already
been noted in some specific contexts.

There are several canonical criticisms which are leveled at
the sort of construction I have used here, and in particular,
at PW time.  Here I give my response to each:

\begin{enumerate}
\item
{\bf Criticism:} PW does not give a good definition of
time because PW time is
not defined in the absence of  subsystems which
behave like good classical clocks.

{\bf Response:}  I am quite happy to let time be a derived
notion which has no meaning in the absence of classical clocks.

\item {\bf Criticism:} Defining probability for arbitrary
states in superspace (as I have) leads to normalization problems.
The infinite extent of time must correspond to an
undboundedness of superspace, making states in
superspace non-normalizable.

{\bf Response:} Since I take time to be a derived notion,
which only makes sense in the presence of classical clocks,
I view the idea that time is infinite with suspicion.
Even a very open universe will have some probability of
experiencing a quantum fluctuation which closes it,
causing classical time to have finite extent.

\item {\bf Criticism:}
Kucha\v{r} \cite{kuchar92} claims that one cannot construct a two-time
correlation function in the  PW picture.

{\bf  Response:} Two time correlation functions {\em can} be
constructed if
care is taken to make an ``operational'' construction which
follows a realistic laboratory procedure as closely as possible.  The
key here is to compare records of observations made at different times
rather than trying to project the poor clock on two different
eigenstates (which of course gives zero).

\end{enumerate}

One possible interpretation of the result form
the previous subsection is that no
matter how carefully one wishes to construct one's theory of the world
out of only the known physics, the theory is
bound to have vast ambiguities.
Even if one constructs a wavefunction describing only what
we know about the
world, evolving according to the observed laws of physics, there will
be other partitions which describe something completely different.

The problem  seems to be that the picture of a state developing in
time is so far removed from a ``state in superspace'' that the two
seem to have very little to do with one another.  If a state
developing in time can be interpreted as a state in superspace, than
it can be interpreted as just about anything.

Another possible
interpretation of the above is that I am being too loose in the way I
go over to the ``state in superspace''.  For one, I am assuming
that there is an inner product defined on $S$ which allows me to
perform all the manipulations required to go from one partition to
another (for example, changing basis from $\left\{
\ketss{k}{S}{}\right\}$ to $\left\{ \ketss{k'}{S}{} \right\}$).
The ambiguities I mention here may be reason enough to avoid
defining such a general inner product on $S$.
On the other hand, just constructing a theory of the universe in the
most conservative manner may already implicitly define a measure on $S$
which is suitable for all the above manipulations.

I hope the analysis presented above will form the basis for further
discussions.  There are  issues at stake which have a
great impact on how one tries to make sense of a
quantum theory of the universe.

\section{Conditional Probabilities}

\subsection{General Remarks}

In the previous section I  argued that processes with which we observe
the world are so ``superficial'' relative to a TOE that uniquely
specifying a TOE may impose very little constraint on the actual
dynamics we observe.  (Of course that raises the question of what is
actually superficial, our observations or the TOE!)

The situation in which a theory presents the scientist with a set of
possible alternatives (rather than a unique prediction) is not unusual
in physics.  A mundane example is a slowly cooled ferromagnet (in the
absence of macroscopic magnetic fields).  No prediction can
be made of the final direction of the magnetization.  None the less,
a lot {\em can } be predicted about this system. In particular at late
times one can predict that the individual spins are highly correlated,
regardless of the overall ambiguity in the direction.

The correlation are exposed by asking ``conditional probability''
questions.  Given that the spin  is ``up'' in one corner of the
magnet, one can predict that elsewhere the spins are  highly likely to
be pointing in the same direction.  In this way one can designate
certain observation to be used as ``conditions'' (e.g. the
magnetization in the corner) and check the remaining observations (the
remaining spin directions) against the predictions.

In the end, physics is about ``sacrificing'' the minimum number of
observations for use as conditions, and
using these conditions to make the
maximum number of predictions.

I would like to advocate that the field of cosmology take this
perspective more seriously.  What are the minimum set of cosmological
observations we should use as `` conditions'' in order to maximize our
predictive power?  I believe that the {\em only} guiding principle in
this process is that minimum number of conditions should be used to
maximum effect.  Questions which can not be resolved even when all the
observations are used as conditions must be classified as metaphysical
questions.

Most cosmologists strongly dislike ``anthropic'' arguments which seem
to be  willing to use too many observations as conditions, and make too
few predictions.
Our job is certainly to get as far  away
{ }from that limit as possible.
It is worth remembering, however, that
quantum mechanics ensures that the generic predictions will take the
form of a probability distribution rather than a single point in
classical phase space, so (as in the ferromagnet example) more
conditions are required to produce concrete predictions.

It is particularly interesting to consider the  implications of the
discussion in Sect. 2 in this context.  If quantum cosmology
provides us with a range of possible effective dynamics, are the laws
of physics at, say, the GUT scale simply a question of metaphysics?
Might different domains with different effective
GUT theories contribute to
the physical processes we observe today?  In Sect. 4 I
explore the implications of this question  on inflationary
cosmologies.

\subsection{The ``Classical'' Condition}
Before considering inflation, I wish to point to a possibly interesting
``condition'' at our
disposal.  Since the invention of quantum mechanics there has been a
lot of concern focused on the extent to which classical behavior
can emerge in quantum systems in the appropriate cases.  Particular
attention has been given to the ``quantum measurement problem''.

I am convinced that as long as the measurement apparatus
is included in the Hilbert space, and one solves the Schr\"{o}dinger
equation for the whole works, there is  no quantum measurement
problem\cite{a93q}.  Even those who oppose this view would agree
that the Schr\"{o}dinger evolution of a realistic apparatus
has very special behavior which does not reflect the properties
of generic quantum systems.
Given this fact, it can seem rather amazing that we are
accustomed to  viewing  this fundamentally quantum world in such a
classical manner.

It might be interesting to turn this question around, and let the
classicality of the universe we observe be one of the {\em conditions} we
use in cosmology (for a related line of thinking see \cite{g-m&h90}).
Certainly very special initial conditions and
special dynamics are required to give classical behavior.  For
example, one
key feature  required for classical behavior is a very clear
statistical arrow of time\cite{a93q}.  This has its roots in
the very low entropy nature of our initial conditions.

It is worth
noting that, as emphasized by Penrose\cite{penrose79},  the sense in
which the early
universe had low entropy is the sense in which the spacetime was very
simple.  It
is these conditions which inflation is so good at producing,
and indeed inflation could well be the {\em best way} of producing
low entropy initial conditions.
Since classical behavior seems to be so atypical of randomly chosen
quantum systems,
the ``classical condition'' may prove to be a very powerful one.
It could (via the arrow of time requirement) end up ``predicting''
inflation, for example.

\section{Implications for Inflation}

Many different effective GUT scale theories
could account for our observations of the physical world.  If quantum
cosmology actually forces us to  consider a range of high
energy physics models as equally realistic alternatives (as I argued
in Sect. 2)  then the implications for inflation
could be very interesting
\footnote{Mezhlumian and Linde give an excellent modern review of the
status inflation with some interesting connections to the issues
raised here.
\cite{mezhlumian&linde94}.
}.

If one asks which inflaton potentials contribute the highest
probability  to finding a universe with whatever feature we choose to
use as  ``conditions'', what type of inflaton potentials are selected
out? It is
quite possible that this perspective gives quite  a different notion
of what is ``natural'' or ``unnatural''  about inflaton potentials.
In particular, if a certain inflaton potential has an unusual
functional form, but  contributes much more
than average to the ``acceptable''
regions of the universe then there is a sense in which this is a
more ``typical'' inflaton potential.

\subsection{An Interesting Possibility}

In an inflationary model the form of the inflaton  potential
determines both the duration of inflation and the spectrum and
amplitude of the perturbations which emerge at the end.  One of the
least attractive features of inflation is that there is no clear
choice for the form of the
inflaton potential\cite{ewk&ma&ejc&arl&jl94,kolb&vadas94}.
Furthermore, all  of
the many candidates look extremely artificial from a particle physics
point of view.  The less artificial versions give unacceptable
predictions (such as density fluctuations so large as to destroy all
the benefits of having inflation in the first place).

Given that all predictions from inflation depend on this  completely
unknown inflaton potential, the theory has essentially no predictive
power.  People try to save inflation from this fate by arguing that
some inflaton potentials  are more ``natural'' or ``generic''.  The
problem is that  by the same criteria the most generic potentials give
unacceptably larger perturbations.

How might the perspective presented in this article change things?
The issues I have raised here may lead to a very different notion of
``typical inflaton potential''.  In my view,  the most attractive
possible outcome would be if inflaton potentials giving
perturbations with  arbitrarily small amplitudes were to make the dominant
contributions (it may be these, for example,  which produce the most
volume of inflated regions).
In this way, any specific details of the potential
would be irrelevant, and the observational predictions of inflation
would be  simple and unique:  A perfectly flat, homogeneous, and
isotropic universe.
Such inflaton potentials would certainly not be ``generic'' by
conventional criteria, but my point is that the conventional criteria
may be the wrong criteria.

The perturbations we observe can then be generated by topological
defects.  This picture would provide
a simple explanation for why the spacetime took  a specific
(unperturbed) form
{\em before } the defects formed.  The lack of such an explanation is
the greatest weakness of the standard defect
scenarios.  The usual attempts to
combine defect scenarios with an earlier period of inflation look even
more ungainly than the standard inflationary picture.

\subsection{Realistic Assessment}

How realistic is the ``zero perturbation'' inflationary scenario
which I  have just described?
The simplest chaotic inflaton potentials
(say $\lambda \phi^4$)  do produce
more expansion (more total volume) as the potential is tuned to
give smaller amplitude perturbations.  However, in that limit the
matter density of the universe at the end of inflation is zero.
Clearly any reasonable ``conditions'' will exclude this possibility.
An attractive condition is that there is net baryon number.  In models
where this can only be generated at the GUT scale there will then be
reheating to temperatures  high enough to produce interesting defects.
(In the very simplest models this may also correspond to
non-negligible perturbations from  inflation,  however.)

Allowing the inflaton potential to take on other than the simplest
forms may actually
decouple the amplitude of the perturbations from the total  volume of
the inflated region.  It may well turn out that these arguments encounter
the same measure problems which appear in the standard discussions of
inflation (what is the right measure to place on the space of inflaton
potentials?)

It is also possible that the measure problem {\em can } be resolved,
but the predictions simply do not coincide with the ``zero
perturbation'' case I am advocating. Another interesting possibility is
that the most sensible ``conditions'' may end up causing inflation to
predict  something really strange,
like $\Omega \neq 1$ \cite{mezhlumian&linde94}.

My conclusion, at this stage, is that these are an interesting set of
ideas which require further investigation before anything concrete can
be built out of them.

\section{Conclusions}

Does it make any sense to discuss an absolutely fundamental TOE?  I
have suggested that even if one could construct
such a theory, the observations we make are so superficial that the
TOE might not constrain these observations in a very substantial way.
In the limit where there is no constraint, a TOE becomes,
{ }from our perspective, a ``Theory of Anything''.

With so much room for ambiguity, one must be absolutely clear about
the boundary between physics and metaphysics.  One important tool in
this effort is the posing of carefully stated conditional probability
questions.  We must state very  carefully what we are
conditioning on and what are we  trying to
predict.  I have argued that the ``classicality'' condition  may prove
to be a very powerful one.

The zero-perturbation inflationary scenario which I have advocated
here seems to suggest an elegant boundary between metaphysics and
physics.  In this scenario the ugly aspects of inflation (the details of
the inflaton potential) are relegated to metaphysics, because the
specific form of the spectrum of perturbations is unobservable.
Whether this attractive scenario can be realized remains to be seen.

A clearer appreciation of the boundary between physics and metaphysics
may prove very sobering for the field of cosmology.  I will end,
therefore,  with an optimistic note:  The field of cosmology has a
grand history of pushing back the boundary between physics and
metaphysics.
Decades ago, who would have thought that the
origin of relative abundances of the different chemical elements would be
considered the subject of physical calculation rather than
metaphysical speculation!  I have no doubt that we will find more
opportunities to push the boundary back even further.

I would like to thank Franco Occhionero for organizing a stimulating
workshop,
and I thank Neil Turok and Robert
Brandenberger for useful conversations.
I am also grateful for the hospitality of the Isaac Newton
Institute where this manuscript was written.

After the workshop, a very
interesting preprint appeared by A. Vilenkin  which has considerable
overlap with the ideas presented here\cite{vilenkin94}.  I thank A.
Vilenkin for discussions of his work.

%\end{document}
%\end

\end{document}